\documentclass[aps,prl,twocolumn,amssymb,nofootinbib]{revtex4}
\pdfoutput=1
\usepackage{epsfig}
\usepackage{amsmath}
\usepackage{amssymb}
\usepackage{graphicx}

\begin{document}
\title{Transient resonances in the inspirals
of point particles into black holes}
\author{ \'Eanna \'E. Flanagan}
\affiliation{Center for Radiophysics and Space Research, Cornell
University, Ithaca, NY 14853, USA}
\author{Tanja Hinderer}
\affiliation{Theoretical Astrophysics, California Institute of Technology, Pasadena, CA 91125, USA}

%\date{\today}
\def\no{\noindent}

\begin{abstract}  We show that transient resonances occur in the
two body problem in general relativity,
for spinning black holes in close proximity to one another when one
black hole is much more massive than the other.
%in the highly relativistic, extreme mass-ratio regime for spinning black holes.
These resonances occur
when the ratio of polar and radial orbital frequencies, which is
slowly evolving under the influence of gravitational radiation
reaction, passes through a low order rational number.
At such points, the adiabatic approximation to the orbital evolution
breaks down, and there is a brief but order unity correction to the
inspiral rate.
%Corrections to the
%gravitational wave signal's phase due to resonance effects scale as the% square root
%of the inverse of mass of the small body, and thus become large in the
%extreme-mass-ratio limit, dominating over all other post-adiabatic effects.
The resonances cause a perturbation to orbital phase
of order a few tens of cycles for mass ratios  $\sim 10^{-6}$,
make orbits more sensitive to changes in initial data (though
not quite chaotic), and are genuine non-perturbative effects that
are not seen at any order in a standard post-Newtonian expansion.
Our results apply to an important potential source of gravitational
waves, the gradual inspiral of white dwarfs, neutron stars, or black
holes into much more massive black holes.
Resonances effects will increase the computational challenge of
accurately modeling these sources.

\end{abstract}

\maketitle

\def\be{\begin{equation}}
\def\ee{\end{equation}}
\def\bfx{{\bf x}}
\def\bfq{{\bf q}}
\def\bfJ{{\bf J}}
\def\bfk{{\bf k}}
\def\bfv{{\bf v}}
\def\bfzero{{\bf 0}}
\def\bfPsi{\mbox{\boldmath $\Psi$}}
\def\bfpsi{\mbox{\boldmath $\psi$}}
\def\bfnabla{\mbox{\boldmath $\nabla$}}
\def\bfsigma{\mbox{\boldmath $\sigma$}}
\def\bfomega{\mbox{\boldmath $\omega$}}
\def\bfOmega{\mbox{\boldmath $\Omega$}}
\def\bfTheta{\mbox{\boldmath $\Theta$}}
\def\bfcalJ{\mbox{\boldmath ${\cal J}$}}
\def\bea{\begin{eqnarray}}
\def\eea{\end{eqnarray}}
\def\nn{\nonumber}
\def\tt{{\tilde t}}
\newcommand{\bes}{\begin{subequations}}
\newcommand{\ees}{\end{subequations}}

\par\noindent{\it Introduction:}
The dynamics of a two-body system emitting gravitational radiation is an
important problem in general relativity.  Binary
systems of compact bodies undergo a radiation-reaction-driven inspiral
until they merge.
There are three different regimes in the parameter space of these
systems: (i) The {\it weak field, Newtonian regime} $r \gg 1$, where
$r$ is the orbital separation in units where $G = c = M=1$ and
$M$ is the total mass.  This regime
can be accurately modeled using post-Newtonian theory, which consists
of an expansion in the small parameter
$1/r$ \cite{PN}. (ii) The {\it relativistic, equal mass
regime} $r \sim 1$, $\varepsilon \sim 1$ (where $\varepsilon = \mu/M$
is the mass ratio with $\mu$ being the reduced mass),
which must be treated using numerical relativity.  Numerical
relativists have recently succeeded for the first time in simulating binary black hole mergers \cite{Pretorius:2007nq}.
%merger of black hole
%binaries, see, for example, the review \cite{Pretorius:2007nq} and
%references therein.
(iii) The {\it relativistic, extreme mass ratio regime}
$r \sim 1$, $\varepsilon \ll 1$, which is characterized by long,
gradual inspirals on a timescale $\sim \varepsilon^{-1}$,
and for which computational methods are currently under development.

In this Letter, we show that in the relativistic,
extreme mass ratio regime, there are qualitatively new aspects to the
two-body problem in general relativity, namely the effects of
transient resonances.  While
resonances are a common phenomenon in celestial mechanics when three
or more objects are involved \cite{ssd}, they can occur with just two
objects in
general relativity, due to its nonlinearity.  They
are a non-perturbative effect for highly relativistic sources
%where the velocity of the small object is close to the speed of
%light,
and are not seen at any order in standard post-Newtonian
expansions of inspiral solutions.
Their existence is closely related to the onset of chaotic dynamics,
which has previously been shown to occur in general
relativity in other, cosmological contexts \cite{chaos}.

The resonances have direct
observational relevance: Compact objects ($1 \alt \mu/M_\odot \alt
10$, where $M_\odot$ is the Solar mass) inspiraling into
much larger black holes are expected to be a key source for
gravitational wave detectors.
Advanced LIGO will potentially
observe $3-30$ such events per year, with $50 \alt M/M_\odot \alt
1000$ \cite{Brown2006}, and
%the future space-based detector LISA is expected
future space-based detectors are expected
to detect
such inspirals with $10^4 \alt M/M_\odot \alt 10^7$ out to cosmological distances at a rate of $\sim 50$
per year \cite{pau}.
The observed gravitational wave signal will be rich in
information. For example, one will be able
to extract a map of the
spacetime geometry of the central object and test if it is
consistent with general relativity's predictions for a black hole
\cite{Barack:2006pq,Brown2006}.
Such mapping will require accurate theoretical models of the gravitational
waveforms, which remain phase coherent with the true waveforms to an
accuracy of $\sim 1$ cycle over the large number  $\sim \varepsilon^{-1} \sim M/\mu \sim
10^2-10^6$ of cycles of inspiral.  Over the past decade
there has been a significant research
effort
% within the general relativity community
aimed at providing
such accurate models
%, see, e.g. Ref.\ \cite{2009CQGra..26u3001B}.
\cite{2009CQGra..26u3001B}.
Resonances will complicate this enterprise, as we discuss below.

%\medskip

\no
{\it Method of Analysis:}
Over timescales short compared to the
dephasing time $\sim \varepsilon^{-1/2}$, inspirals can be accurately
modeled using black hole perturbation theory, % \cite{Teukolsky:1973ha},
with $\varepsilon$ as the expansion parameter.
The leading order motion is geodesic motion on the background
Kerr metric. % of the central body.
At the next order
the motion is corrected by the particle's self-force or radiation
reaction force, for which a formal expression is known \cite{Mino:1996nk},
and which has been computed explicitly in special cases; see, e.g.,
the review \cite{2009CQGra..26u3001B}.
Over the longer inspiral timescale $\sim \varepsilon^{-1}$, it is
necessary to augment these methods with two-timescale expansions which are currently
under development \cite{paperI,Minoscalar}.  In this
framework the leading order motion is an adiabatic inspiral,
%which is approximately geodesic at each instant in time,
and there are various
post-adiabatic corrections.

Geodesic motion in the Kerr spacetime is an integrable dynamical
system, and it is useful to use the
the corresponding generalized action-angle variables to parameterize
the inspiral.  The resulting equations are \cite{FH08b}:
\bes
\bea
\label{dqdtau}
\frac{d q_\alpha }{d\tau} &=& {\omega}_\alpha({\bf J})
+\varepsilon
g_\alpha^{(1)}(q_\theta,q_r, {\bf J})  + O(\varepsilon^2),\\
\label{djdtau}
\frac{d J_\nu}{d\tau} &=& \varepsilon
G^{(1)}_\nu(q_\theta,q_r, {\bf J}) + \varepsilon^2
G^{(2)}_\nu(q_\theta,q_r, {\bf J})  +
O(\varepsilon^3).\ \ \ \ \ \
\eea
\label{eq:eomsimplified}
\ees
Here $\tau$ is Mino time \cite{scalar} and $J_\nu$ are the conserved
integrals of
geodesic motion given by
$J_\nu=(E/\mu, L_z/\mu,Q/\mu^2)$, where $E$ is the energy, $L_z$
is the angular momentum, and $Q$ the Carter constant.  The variables
$q_\alpha = (q_t,q_r,q_\theta,q_\phi)$ are
generalized angle variables conjugate to Mino time \cite{paperI}.
%the generalized angle
%variables \cite{paperI} corresponding to motions in the $t,r,\theta$
%and $\phi$ directions in Boyer-Lindquist coordinates.
The right hand sides at $O(\varepsilon^0)$ describe geodesic motion,
with fundamental frequencies $\omega_r$, $\omega_\theta$ and
$\omega_\phi$.  The forcing functions $g^{(1)}_\alpha$,
$G^{(1)}_\nu$ and $G^{(2)}_\nu$ are due to the first order and second
order self-forces, and are $2\pi$-periodic in $q_\theta$ and $q_r$.
%and
%$G^{(1)}_\nu$ are determined by the first order self-force, and
%the function $G^{(2)}_\nu$ by the second order self-force;
%these are $2\pi$-periodic functions of $q_\theta$ and $q_r$.
The piece of $G^{(1,2)}_\nu$ that is even under $q_\theta \to 2 \pi - q_\theta$,
$q_r \to 2 \pi - q_r$, and the piece of $g^{(1)}_\alpha$ that is odd,
are the dissipative self-force, and the remaining
piece is the conservative self-force \cite{paperI}.

In the limit $\varepsilon \to 0$, the leading order solutions to Eqs.\
(\ref{eq:eomsimplified}) are given by the following {\it adiabatic
prescription} \cite{paperI}: Drop the forcing terms
$g^{(1)}_\alpha$ and $G^{(2)}_\nu$, and replace $G^{(1)}_\nu$
by its average over the 2-torus parameterized by $q_\theta,q_r$.
It is now known how to evaluate this averaged force explicitly
\cite{Mino2003,scalar},
although generic adiabatic inspirals have not yet been computed
numerically.

Consider now post-adiabatic effects.  The dynamical system
(\ref{eq:eomsimplified}) consists of a perturbed, integrable
Hamiltonian system.  Resonances in this general type of system
have been studied in detail and are well understood
%, see for example monograph
\cite{Kevorkian}, and we can apply the
general theory to the present context.
The existence of resonances in this
context has previously been suggested by
Refs.\ \cite{2005PThPh.113..733M,haris}.
%In the remainder of this paper
We will present three different
treatments of the resonances: (i) An intuitive, order of magnitude
discussion, which is sufficient to deduce their key properties; (ii) A
numerical treatment; and (iii) A sketch of a formal analytic
derivation.  A more detailed treatment will be presented in Ref.\ \cite{FH08b}.

\medskip
\no
{\it Order of Magnitude Estimates:} Suppose that we have an adiabatic
solution, which will be of the form
$q_\alpha(\tau,\varepsilon) = \psi_\alpha(\varepsilon
\tau)/\varepsilon$, $J_\nu(\tau,\varepsilon) =
J_\nu(\varepsilon\tau)$.  Consider now the post-adiabatic correction
terms in Eqs.\ (\ref{eq:eomsimplified}),
% in the neighborhood of
near some
arbitrarily chosen point $\tau=0$.
We expand $q_\theta$ as $q_\theta = q_{\theta 0} + \omega_{\theta 0} \tau +
{\dot \omega}_{\theta 0} \tau^2 + O(\tau^3)$, where subscripts 0 denote
evaluations at $\tau=0$, and we expand $q_r$ similarly.
We also expand % the forcing function
$G^{(1)}_\nu$ as a double Fourier
series:
%\be
%G^{(1)}_\nu(q_r,q_\theta,
%{\bf J})=\sum^{\infty}_{k_r=-\infty}\sum^{\infty}_{k_\theta=-\infty}G^{%(1)}_{\nu
%~ k_rk_\theta}({\bf J} )e^{i(k_rq_r+k_\theta q_\theta)},
%\nonumber
%\ee
$
G^{(1)}_\nu(q_\theta,q_r,
{\bf J})=\sum_{k,n} G^{(1)}_{\nu
~ k n}({\bf J} )e^{i(k q_\theta+n q_r)},
$
where the $00$ term is the adiabatic approximation,
and the remaining terms drive post-adiabatic effects.  Inserting the
expansions of $q_\theta$ and $q_r$, we find for the phase of the $(k,n)$
Fourier component
\be
({\rm constant})+(k \omega_{\theta0} + n \omega_{r0}) \tau  +
(k {\dot \omega}_{\theta0} + n {\dot \omega}_{r0})
\tau^2  + \ldots
\label{phase}
\ee
Normally, the second term is nonzero and thus the force oscillates on a
timescale $\sim 1$, much shorter than the inspiral timescale $\sim
1/\varepsilon$, and so the force averages to zero.  However, when the
resonance condition
$
k \omega_{\theta0} + n \omega_{r0} =0
$
is satisfied, the $(k,n)$ force is slowly varying and cannot
be neglected, and so gives an order-unity correction to the right hand
side of Eq.\ (\ref{djdtau}).  The duration of the resonance is given
by the third term in (\ref{phase}) to be $\tau_{\rm res} \sim
1/\sqrt{v {\dot \omega}}  \sim 1/\sqrt{v\varepsilon}$, where $v = |k| + |n|$ is the order of the
resonance;
%\footnote{We treat the resonances as isolated but in
%fact there is an infinite number of resonances. However, both
%analytical considerations and numerical simulations show that only
%low order resonances are important in practice
%\protect{\cite{FH08b}}.}
after times longer than this the quadratic term causes the force to
oscillate and again average to zero.
The net change in the action variables $J_\nu$ is therefore
$\Delta J_\mu \sim {\dot J} \tau_{\rm res} \sim \varepsilon \tau_{\rm res} \sim \sqrt{\varepsilon/v}$.
After the resonance, this change causes
a phase error $\Delta \phi$ that accumulates
over an inspiral, of order the total inspiral phase $\sim
1/\varepsilon$ times $\Delta J/J \sim \sqrt{\varepsilon/v}$, which gives
$
\Delta \phi \sim 1/\sqrt{v \varepsilon}.
$

This discussion allows us to deduce several key properties of the
resonances.  First, corrections to the
gravitational wave signal's phase due to resonance effects scale as the square root
of the inverse of mass of the small body.  These corrections thus
become large in the extreme-mass-ratio limit, dominating over all
other post-adiabatic effects, which scale as $\varepsilon^0 \sim 1$.

  Second, they occur when $\omega_r/\omega_\theta$ is a low order rational
  number.  There is a simple geometric picture corresponding to this
  condition \cite{2008PhRvD..77j3005L,haris}: the geodesic
  orbits do not ergodically fill out the $(q_\theta,q_r)$ torus in space
  as generic geodesics orbits do, but instead form a 1 dimensional curve on the
  torus.  This implies that the time-averaged forces for these orbits
  are not given by an average over the torus, unlike the case for generic orbits.

Third, they occur only for non-circular, non-equatorial orbits about
  spinning black holes.  For other cases, the forcing terms
  $G^{(1)}_\nu$ depend only on $q_\theta$, or only on $q_r$, but not
  both together, and thus the Fourier coefficient $G^{(1)}_{\nu\,kn}$ will vanish for any resonance.

Fourth, they are driven only by the spin-dependent part of the
  self-force, for the same reason:
%  The $O(a^0)$ piece is the self-force in
%Schwarzschild spacetime, for which spherical symmetry forbids a
%dependence on $q_\theta$.
spherical symmetry
forbids a dependence on $q_\theta$ in the zero-spin limit.

Fifth, they appear to be driven only by the dissipative part of the
  self-force, and not by the conservative part, again because the
  forcing terms do not depend on both $q_\theta$ and $q_r$.
We have verified that
this is the case up to the post-Newtonian order that
spin-dependent terms have been computed \cite{2006PhRvD..74j4033F},
and we conjecture that it is true to all orders.  The reason that this
occurs is that the conservative sector of post-Newtonian theory admits
three independent conserved angular momentum components; the
ambiguities in the definition of angular momentum are associated with
radiation, in the dissipative sector.  As a consequence, the perturbed
conservative motion is integrable to leading order in $\varepsilon$, and an
integrable perturbation to a Hamiltonian cannot drive resonances.

Sixth, although the resonance is directly driven only by dissipative,
  spin dependent self-force, computing resonance effects requires
the conservative piece of the first order self-force and
the averaged, dissipative piece of the second order self-force.
Those pieces will cause
$O(1)$ corrections to the phases over a complete inspiral
\cite{paperI}, and the kicks $\Delta J_\mu$ produced during the
resonance depend on the $O(1)$ phases at the start of the resonance.

Seventh, resonances give rise to increased sensitivity to initial conditions,
analogous to chaos but not as extreme as chaos, because at a resonance
information flows
from a higher to a lower order in the perturbation expansion.  For example,
we have argued that changes to the phases at $O(1)$ prior to the resonance will affect
the post-resonance phasing at $O(1/\sqrt{\varepsilon})$.  Similarly
changes to the phases at $O(\sqrt{\varepsilon})$ before resonance will
produce $O(1)$ changes afterwards.  With several successive resonances,
a sensitive dependence on initial conditions could arise.

\begin{figure}
\begin{center}
%{\includegraphics[scale=0.6]{2a}}
%{\includegraphics[scale=0.6]{2b}}
\includegraphics[width=8.5cm]{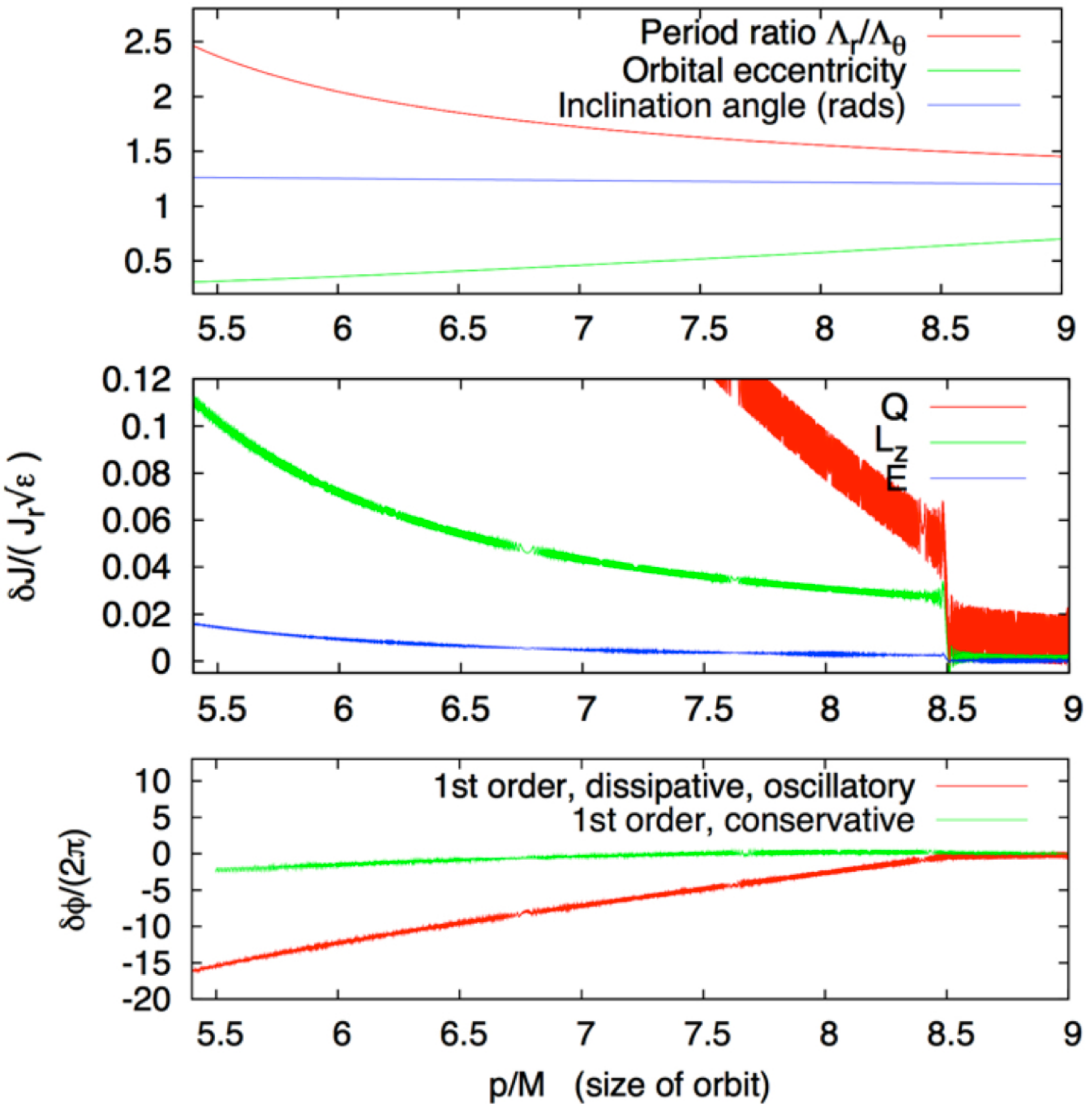}
\caption{[Top] The adiabatic inspiral computed
from our approximate post-Newtonian self-force, for a mass ratio
$\varepsilon = \mu/M = 3 \times 10^{-6}$, with black hole spin parameter $a = 0.95$, with initial
conditions semilatus rectum $p =9.0 M$, eccentricity $e = 0.7$, and
orbital inclination $\theta_{\rm inc} = 1.20$.  The bottom curve is
$e$, the middle curve is $\theta_{\rm inc}$, and the top curve is
ratio of frequencies
$\omega_\theta/\omega_r$, shown as functions of $p$.
[Middle] The fluctuating, dissipative part of the first order
self-force
% gives rise to corrections to the adiabatic approximation, and in
% particular
causes a strong resonance
when $\omega_\theta/\omega_r = 3/2$ at $p = 8.495$.
Shown are the corrections to the energy $E$, angular momentum $L_z$ and Carter constant $Q$, as functions of $p$, scaled to their values at resonance, and divided by the square root $\sqrt{\mu/M}$ of the mass ratio.  The sudden jumps at the resonance are apparent, with the largest occurring for the Carter constant.
[Bottom] The lower curve is the correction to the number of cycles
$\phi/(2 \pi)$ of azimuthal phase of the inspiral caused by the
fluctuating, dissipative part of the first order self-force.
The sharp
downward kick due to the resonance at $p = 8.495$ can be clearly seen.  The resonant corrections to the
number of cycles of $r$ and $\theta$ motion are similar.  These phase
shifts scale as $\sqrt{M/\mu}$.
The upper curve is the post-adiabatic phase correction due to the
conservative piece of
the first order self-force, which is considerably smaller and
is independent of the mass ratio.}
\label{fig1}
\end{center}
\end{figure}

\medskip
\no
{\it Numerical Integrations:} The scaling relation
$ \Delta \phi \propto 1/\sqrt{\varepsilon}$
% for the phase errors produced by resonances
suggests the possibility of phase errors large compared to
unity that impede the detection of the gravitational wave signal.
To investigate this possibility, %in more detail
we numerically integrated the
exact Kerr geodesic equations supplemented with approximate
post-Newtonian forcing terms.  While several
such approximate inspirals have been computed previously
\cite{Cutler:1994pb},
none have encountered resonances, because resonances require
non-circular, non-equatorial orbits about a spinning black hole with
non orbit-averaged forces, which have not been
simulated before.

%It is not convenient to use the angle variables $q_\alpha$ for
%numerics, since they are not known analytically in closed form.
%Instead we use the variables
For the numerical integrations we use instead of $q_\alpha$ the variables
${\bar q}_\alpha = ({\bar q}_t,{\bar q}_r, {\bar q}_\theta, {\bar
  q}_\phi) = (t, \psi,\chi,\phi)$ where  $\psi$ and $\chi$ are the
angular variables for $r$ and $\theta$ motion defined in Ref.\
\cite{scalar}.
The equations of motion (\ref{eq:eomsimplified}) in these variables are
\bes
\bea
t_{,\tau}
 &=& {\bar \omega}_{t}({\bar q}_\theta,{\bar q}_r,{\bf J}),  \ \
\phi_{,\tau} = {\bar \omega}_{\phi}({\bar q}_\theta,{\bar q}_r,{\bf J}), \\
{\bar q}_{\theta,\tau} &=& {\bar \omega}_\theta({\bar q}_\theta,{\bf J}) + \varepsilon h^{(1)}_\theta({\bar q}_\theta,{\bar q}_r,{\bf J}) + O(\varepsilon^2),\\
{\bar q}_{r,\tau} &=& {\bar \omega}_r({\bar q}_r,{\bf J}) + \varepsilon h^{(1)}_r({\bar q}_\theta,{\bar q}_r,{\bf J}) + O(\varepsilon^2),\\
J_{\nu,\tau} &=& \varepsilon H^{(1)}_\nu({\bar q}_\theta,{\bar q}_r,{\bf J}) + O(\varepsilon^2).
\eea
\label{eq:numeric}
\ees
Here $\tau$ is Mino time \cite{scalar}, the frequencies
${\bar \omega}$ are given in \cite{scalar}, and
$h^{(1)}_\alpha$ and $H^{(1)}_\nu$ are given in terms of the components of the
4-acceleration in \cite{Skypers}.

We parameterize the three independent components of the acceleration in the following way:
$
a^\alpha = a^{{\hat r}} e_{{\hat r}}^\alpha + a^{{\hat \theta}} e_{{\hat \theta}}^\alpha + a_\perp \epsilon^{\alpha}_{\ \beta\gamma\delta} u^\beta e^\gamma_{{\hat r}} e^\delta_{{\hat\theta}} + (a^{{\hat r}} u_{{\hat r}} + a^{{\hat \theta}} u_{{\hat \theta}}) u^\alpha,
$
where ${\vec u}$ is the 4-velocity and ${\vec e}_{\hat r}$ and ${\vec
  e}_{\hat \theta}$ are unit
vectors in the directions of $\partial_r$ and $\partial_\theta$.
We compute the dissipative pieces of $a^{{\hat r}}$, $a^{{\hat
    \theta}}$ and $a_\perp$ from the results of
\cite{2007PhRvD..75l4007F}, as functions of
${\tilde r} = r + a^2/(4 r)$, $E_n = E-1$, and ${\bar K} = Q + a^2
L_z^2 + a^2E_n$, and then expand to $O(a^2)$ and to the
leading post-Newtonian order at each order in $a$ \cite{FH08b}.
We also add the conservative component, expressed similarly and
computed to $O(a)$ and to the leading post-Newtonian order, taken from
Refs.\ \cite{1993PhRvL..70..113I};
see Ref.\ \cite{FH08b} for details.
%Full details of this computation will be presented elsewhere.

We numerically integrate Eqs.\ (\ref{eq:numeric}) twice, once using the
adiabatic prescription, and once exactly, and then subtract at fixed
$t$ to obtain the post-adiabatic effects.  The adiabatic prescription
involves numerically integrating the right hand sides over the torus
parameterized by $q_\theta,q_r$ at each time step, where
%$$
%q_r = \int_0^{{\bar q}_r} \frac{d{\bar q}_r }{ {\bar \omega}_r({\bar
%    q}_r,{\bf J})} /
%\int_0^{2\pi} \frac{d{\bar q}_r }{ {\bar \omega}_r({\bar
%    q}_r,{\bf J})},
%$$
$q_r = F_r({\bar q}_r)/F_r(2 \pi)$,
$F_r({\bar q}_r) = \int_0^{{\bar q}_r} d{\bar q}_r / [{\bar \omega}_r({\bar
    q}_r,{\bf J})],
$
with a similar formula for $q_\theta$.  This is numerically time
consuming, but the adiabatic integration can take timesteps on the
inspiral timescale $\sim 1/\varepsilon$ rather than the dynamical
timescale $\sim 1$.

Typical results are shown in Fig.\ \ref{fig1}, which shows the
adiabatic inspiral for a mass ratio of $\varepsilon = 3 \times
10^{-6}$ with $a = 0.95$,
in terms of the relativistic eccentricity $e$,
semilatus rectum $p$ and orbital inclination $\theta_{\rm inc}$,
which are functions of $E$, $L_z$ and $Q$ \cite{Drasco:2005kz}.  This
example has a strong resonance at
$\omega_\theta/\omega_r = 3/2$, that generates jumps in the conserved
quantities of order a few percent times $\sqrt{\varepsilon}$, and
causes phase errors over the inspiral of order 20 cycles.
Phase errors of this magnitude
% This large
%of a phase error
will be a significant impediment to signal detection
with matched filtering.
We find that the resonance effects are dominated
by the $O(a^2)$ terms, and the effect of the $O(a)$ terms are small.

\medskip
\no
{\it Analytic Derivation:} In terms of the slow time variable
$\tilde \tau=\varepsilon \tau$, the solutions of the dynamical system
(\ref{eq:eomsimplified}) away from resonances can be expressed as an
asymptotic expansion in $\varepsilon$ at fixed ${\tilde \tau}$
  \cite{paperI,Kevorkian}:
\bes
\label{asymptsol}
\bea
\label{phasing}
q_\alpha(\tau,\varepsilon)&=&\frac{1}{\varepsilon} \left[
\psi^{(0)}_\alpha(\tilde \tau)
+\sqrt{\varepsilon}\psi^{(1/2)}_\alpha(\tilde\tau) + O(\varepsilon) \right],\\
J_\nu(\tau, \varepsilon)&=& {\cal
J}^{(0)}_\nu(\tilde \tau)+\sqrt{\varepsilon}{\cal
J}^{(1/2)}_\nu(\tilde \tau)+O(\varepsilon).
\eea\ees
The leading order terms give the adiabatic approximation described
above, and satisfy \cite{paperI}
$\psi^{(0)}_{\alpha,{\tilde \tau}} = \omega_\alpha[{\bfcalJ}^{(0)}]$,
${\cal J}^{(0)}_{\nu,{\tilde \tau}} =
\langle G^{(1)}_{\nu}\rangle[ \bfcalJ^{(0)}]$,
where the angular brackets denote an average over the $(q_r,q_\theta)$
torus.
The subleading, post-$1/2$-adiabatic order terms satisfy
%\bes\bea\frac{d{\cal
%J}^{(1/2)}_\nu}{d\tilde \tau}-\frac{\partial\langle
%G^{(1)}_\nu\rangle }{\partial J_\mu}{\cal
%J}^{(1/2)}_\mu&=&\Delta J^{(1/2)}_\nu \delta(\tau),\label{Jhalf}\\
% \frac{d{\psi}^{(1/2)}_\alpha}{d\tilde \tau}-\frac{\partial
%\omega_\alpha }{\partial J_\mu}{\cal J}^{(1/2)}_\mu&=&0,
%\label{psihalf}
%\eea \ees
${\cal J}^{(1/2)}_{\nu,{\tilde \tau}}-\langle
G^{(1)}_\nu\rangle_{,J_\mu} {\cal J}^{(1/2)}_\mu=\Delta J^{(1/2)}_\nu \delta(\tau)$,
${\psi}^{(1/2)}_{\alpha,{\tilde \tau}}= \omega_{\alpha,J_\mu}{\cal J}^{(1/2)}_\mu$,
where the $\delta$-function source term arises at a resonance, taken
to occur at $\tau =0$.

Near the resonance we use an ansatz for the solutions which is an
asymptotic expansion in $\sqrt{\varepsilon}$ at fixed ${\hat \tau} =
\sqrt{\varepsilon} \tau$, and then match these solutions onto
pre-resonance and post-resonance solutions of the form
(\ref{asymptsol}) \cite{Kevorkian}.
To linear order in the force Fourier coefficients (an approximation
which is valid here to within a few percent \cite{FH08b}), the jumps in the
action variables for a resonance $(k,n)$ can be computed by
substituting the adiabatic
solutions into the right hand side of
Eqs.\ (\ref{eq:eomsimplified}) and solving for the perturbation to the
action variables.  The result is
\be
\Delta J_\nu^{(1/2)} = \sum_{s \ne 0} \sqrt{ \frac{2 \pi}{| \alpha s|
  }} \exp \left[ {\rm sgn}(\alpha s) \frac{i \pi}{4} + i s \chi_{\rm
    res} \right] G^{(1)}_{\nu\,sk,sn},\nonumber
\ee
where $\chi_{\rm res} = k q_\theta + n q_r$, $\alpha = k
\omega_{\theta,{\tilde \tau}} + n \omega_{r,{\tilde \tau}}$ and all
quantities are evaluated at the resonance $\tau=0$ using the adiabatic
solution.  In Ref.\ \cite{FH08b} we give the exact expression for this
quantity that does not linearize in the force Fourier coefficients.
We note that evaluating the phase $\chi_{\rm res}$ requires knowledge
of the second subleading, $O(1)$ phase in Eq.\ (\ref{phasing}), which
in turn requires knowledge of the force components
$g^{(1)}_\alpha$, $G^{(1)}_\nu$ and $\langle G^{(2)}_\nu\rangle$ in
Eq.\ (\ref{eq:eomsimplified}) \cite{paperI}.
In addition, to obtain the phase to $O(1)$ accuracy after the
resonance,
%(needed for initial conditions for any
%subsequent resonances)
it is necessary to also compute the subleading,
$O(\varepsilon)$ jumps in $J_\nu$
%(because changes in
%$J_\nu$ at $O(\varepsilon)$ affect the post-resonance phases at
%$O(1)$ by a similar argument as above)
and $O(1)$ jumps in $q_\alpha$, which
are given in \cite{FH08b}.

\medskip

\no
{\it Discussion:} The dynamics of binary systems in general relativity
is richer than had been appreciated.  Transient resonances occurring
during the inspiral invalidate the adiabatic approximation and give
rise to corrections to the orbital phase that can be large compared to
unity.  It will be necessary to incorporate resonances into
theoretical models of the gravitational waveforms for inspirals of
compact objects into massive black holes, an important gravitational
wave source.
This will require knowledge of
the second order gravitational self force and will be challenging.

\medskip

\no
{\it Acknowledgments:} This work was supported by NSF
Grants PHY-0757735 and PHY-0457200, by
the John and David Boochever Prize Fellowship in Theoretical Physics to TH at Cornell,
and by the Sherman Fairchild Foundation.
We thank Scott Hughes, Marc Favata, Steve Drasco and Amos Ori for
helpful conversations.
%EF thanks the Theoretical Astrophysics Including Relativity Group at Caltech, and the Department of Applied Mathematics and Theoretical Physics at the University of Cambridge, for their hospitality as this paper was being completed.

%\bibliographystyle{apsrev}
%\bibliographystyle{physrev}
%\bibliographystyle{prl}
%\bibliography{refs}

\begin{thebibliography}{47}

\bibitem{PN}
L.~Blanchet, Living Reviews in Relativity \textbf{5}, 3 (2002).

\bibitem{Pretorius:2007nq}
F.~Pretorius (2007), arXiv:0710.1338.

\bibitem{ssd}
C.~D.~Murray and S.~F.~Dermott,
\emph{Solar System Dynamics}
(Cambridge University Press, 2000).

\bibitem{chaos}
N.J.\ Cornish and J.\ Levin, Phys. Rev. Lett. \textbf{78}, 998 (1997);
J.D.\ Barrow and J.\ Levin, Phys. Rev. Lett. \textbf{80}, 656 (1998).

\bibitem{Brown2006}
D.~A.\ Brown et~al., Phys. Rev. Lett. \textbf{99}, 201102 (2007);
I.~Mandel et.\ al., \apj \textbf{681}, 1431 (2008).

\bibitem{pau}
P.\ Amaro-Seoane et~al., Class. Quant. Grav. \textbf{24}, R113 (2007).

\bibitem{Barack:2006pq}
L.~Barack and C.~Cutler, Phys. Rev. D \textbf{75},042003 (2007).

%\bibitem{capra}
%Presentations at the meeting "Theory Meets Data Analysis at
%Comparable and Extreme Mass Ratios", Perimeter Institute, Canada,
%June 2010, available at http://www.perimeterinstitute.ca/Events/.

\bibitem{2009CQGra..26u3001B}
L.~Barack, Class. Quant. Grav. \textbf{26}, 213001 (2009).

%\bibitem{Teukolsky:1973ha}
%S.~A. Teukolsky, Astrophys. J. \textbf{185}, 635 (1973).

\bibitem{Mino:1996nk}
Y.~Mino et al.,
%Y.~Mino,M.~Sasaki and T.~Tanaka,
Phys. Rev. \textbf{D55}, 3457 (1997);
T.~C.~Quinn and R.~M. Wald, Phys. Rev. \textbf{D56},
3381 (1997).
%; S.~E. Gralla and R.~M. Wald, Class. Quant. Grav.
%\textbf{25}, 205009 (2008).

%\bibitem{Barack:2010tm}
%L.~Barack and N.~Sago,
%Phys. Rev. \textbf{D81},
%084021 (2010); T.~S. Keidl et al.,ArXiv:1004.2276; S.~Detweiler, \prd
%\textbf{77}, 124026 (2008);
% R.~Haas and E.~Poisson,
%Phys. Rev. \textbf{D74}, 044009 (2006);
%P.~Ca\~nizares and C.~F. Sopuerta, \prd \textbf{79}, 084020 (2009);
%N.~Warburton and L.~Barack, \prd \textbf{81},
%084039 (2010).

\bibitem{paperI}
T.~Hinderer and \'E.~\'E.\ Flanagan,
\prd \textbf{78},
064028 (2008).

\bibitem{Minoscalar}
Y.~Mino and R.~Price,
Phys. Rev. \textbf{D77}, 064001 (2008);
A.~Pound and E.~Poisson, Phys. Rev. \textbf{D77}, 044012 (2008).

%\bibitem{FH08c}
%E.~E. Flanagan and
%T.~Hinderer,
%in preparation  (2010).

\bibitem{Mino2003}
Y.~Mino, Phys.
Rev. \textbf{D67}, 084027
(2003); S.~A. Hughes et al.,
Phys. Rev. Lett. \textbf{94},
221101 (2005);
%  N.~Sago et al.,
%Progress of Theoretical Physics
%\textbf{114}, 509 (2005);
N.~Sago et al., Prog. Theor.
Phys. \textbf{115}, 873 (2006);
%T.~Hinderer, Ph.D. thesis,
%Cornell (2008);
R.~Fujita et al.,
Prog. Theor. Phys.
\textbf{121}, 843 (2009).

\bibitem{scalar}
S.~Drasco et al., Class. Quant. Grav.
\textbf{22}, S801 (2005).

\bibitem{Kevorkian}
J.~Kevorkian and
J.~D. Cole,
{\it Multiple scale and singular perturbation methods}
Springer, New York, 1996.

\bibitem{2005PThPh.113..733M}
Y.~Mino,
Prog. Theor. Phys.
\textbf{113}, 733 (2005);
T.~Tanaka,
Prog. Theor. Phys. Suppl. \textbf{163},
120 (2006).

\bibitem{haris}
T.~A. Apostolatos et al.,
Phys. Rev. Lett. \textbf{103},
111101 (2009).

\bibitem{FH08b}
E.~E. Flanagan and
T.~Hinderer, in preparation.

\bibitem{2008PhRvD..77j3005L}
J.~Levin and
  G.~Perez-Giz,
  \prd \textbf{77},
  103005 (2008).

\bibitem{2006PhRvD..74j4033F}
G.~Faye et al.,
\prd \textbf{74},
104033 (2006).

\bibitem{Cutler:1994pb}
C.~Cutler et al.,
Phys. Rev. D \textbf{50},
3816 (1994); S.~A.Hughes,
Phys. Rev. D \textbf{61},
084004 (2000);
S.~Babak et al., Phys. Rev. D
\textbf{75}, 024005 (2007);
J.~R. Gair and
K.~Glampedakis,
\prd \textbf{73},
064037 (2006);
N.~Yunes et al.,
Phys. Rev. Lett. \textbf{104},
091102 (2010).

\bibitem{Skypers}
J.~Gair et al., Phys. Rev. D {\bf 83}, 044037 (2011).

\bibitem{2007PhRvD..75l4007F}
\'E.~\'E. Flanagan
and
T.~Hinderer,
\prd \textbf{75},
124007 (2007).

\bibitem{1993PhRvL..70..113I}
B.~R. Iyer and
C.~M. Will,
Phys. Rev. Lett. \textbf{70},
113 (1993);
L.~E. Kidder,\prd \textbf{52}, 821
  (1995).

\bibitem{Drasco:2005kz}
S.~Drasco and
S.~A. Hughes\,
Phys. Rev. D \textbf{73},
024027 (2006).

\end{thebibliography}

%%% Local Variables:
%%% mode: latex
%%% TeX-master: t
%%% End:

\end{document}